\documentclass[pre,reprint,twocolumn]{revtex4}
\pdfoutput=1 
\usepackage{graphicx}
\usepackage{color}
\usepackage{amsmath}
\usepackage{amssymb}

\begin{document}

\title{On the existence of an upper critical dimension for systems within the KPZ universality class}

\author{Evandro A. Rodrigues }
\author{Fernando A. Oliveira}
\email{faooliveira@gmail.com}
\author{Bernardo A. Mello }
\affiliation{Institute Of Physics, University of Brasilia, Caixa Postal 04513, 70919-970 Brasilia-DF, Brazil}

\begin{abstract}

In this work we extend the etching model ~\cite{Mello2001} to $d+1$ dimensions. This permits us to investigate
its exponents behaviour on higher dimensions, to try to verify the existence of an upper critical dimension
for the KPZ equations, with our results sugesting that  $d=4$ is  not an upper critical dimension for the etching model.

\end{abstract}


\date{\today}
\maketitle
\begin{section}{Introduction}

Among the many equations describing the behaviour of moving surfaces, the KPZ equation ~\cite{Kardar1986}  is of
special interest, partially motivated by the fact that it describes many physical phenomena, such as flame front 
propagation \cite{Provatas1995,Campos2013} and deposition of thin films \cite{Lita2000}.
Although its exponents in one dimension are long known ~\cite{Hwa1991,Frey1996}, the renormalization
group technique used for obtaining its values is not usable for higher dimensions.
As such, usualy either numerical simulations or approximate methods are used to obtain exponents for the KPZ in higher dimensions, 
and even then numerical determination of exponents can be tricky~\cite{Wio2013}.

Such methods give no final answer to the values of these exponents, which combined with the
absence of exact solutions leads to the much debated possibility of an
upper critical dimension (UCD) $d_c$ for the dynamic exponents\cite{Katzav2002}.

One approach consisted of trying to find a general expression for the exponents
depending on the substrate dimension $d$, i.e. $\alpha=\alpha(d)$, 
$\beta=\beta(d)$ and $z=z(d)$. 
Notable examples  are those for the RSOS model,  by Kim and 
Kosterlitz~\cite{Kim1991}, for the Eden model \cite{Moser1991} by  Kert\'esz and Wolf,  
the heuristic approach to the strong-coupling regime by Stepanow \cite{Stepanow1997},
a tentative method based on  quantization of the exponents by  L\"assig\cite{Lassig1998} and
a pertubation expansion of the KPZ equation by Bouchaud and Cates \cite{Bouchaud1993}.
Unfortunately, further numerical results have shown these results to
be incorrect~\cite{Forrest1990,Tang1992,Ala-Nissila1993,Ala-Nissila1998}.

Analytical methods such as mapping of the directed polymer \cite{Halpin-Healy1989}, 
perturbation expansion \cite{Wiese1998} and mode-coupling techniques
\cite{Bhattacharjee1998} among others
\cite{Lassig1995,Lassig1997,T.Blum1995,M.A.MooreT.BlumJ.P.Doherty1995} observe $d_c = 4$.
A asymptotic weak noise approach by Fogedby  \cite{Fogedby2009} suggests $d_c < 4$.
Otherwise, some numerical studies found no such limit
\cite{Moser1991,Marinari2000,Marinari2002}, 
as well as the numerical and theoretical results by Scharwartz and Perlsman \cite{Schwartz2012}.

We present in this work a version of the Etching model by Mello et al~\cite{Mello2001,Mello2015}
extended to work with $d+1$ spatial dimensions~\cite{Rodrigues2015}. The one dimensional version of the
Etching model is known to be compatible with the KPZ equation, and as such,
is classified in the KPZ universality class.

Using this version of the model, we determine exponents for surfaces with $1\le d\le 6$
, reaching the conclusion that if
there is a UCD, it must be such that $d_c > 6$.

\end{section}
\begin{section}{The etching model in $d$ dimensions}

Proposed by Mello et al~\cite{Mello2001} in 2001, the etching model simulates a 
one dimensional crystalline solid submerged in a solvent liquid. Its scaling exponents
are  very close to those of the KPZ equation, namely $\alpha = 0.4961 \pm 0.0003$
and $\beta = 0.330\pm0.001$, and as such, it is believed do belong to the KPZ universality class.
This model was object of extensive research in recent years~\cite{AaraoReis2006a,Paiva2007,Kimiagar2008,Oliveira2008,Tang2010,Xun2012}.

We extend the model to $d+1$ dimensions, considering the solid a square lattice exposed to
the solvent liquid, with a removal probability proportional to the exposed area. The algorithm 
can be described as
\begin{enumerate}
\item at discrete instant $T$ one horizontal site $i=1,2...,L$ is randomly
    chosen;
\item $h_i(T+1) = h_i(T)+1$;
\item if $h_{i+\delta}(T) < h_i(T)$, do $h_{i+\delta}(T+1) = h_i(T)$, where
$\delta=\pm 1$ are the first neighbours. 
\end{enumerate}
In the multidimensional version $i$ and $\delta$ are vectors and $\delta$ runs over the
$2^d$ first neighbours of the hypercube. We consider $L$ to be the substrate length in each
direction, with the total number of sites is $L^d$. The normalized time $t$ defines
the time unity as $L^d$ cellular automata iterations, i.e., $t=T/L^d$. We use periodic boundary
conditions on the surface to reduce unwanted finite size length effects.
Albeit the model is not a direct mapping of the KPZ equation, it generally mimics its dynamics
and reproduces its exponents for $1+1$ as well as the general case $d+1$.

\begin{figure}[hp!]
\includegraphics[width=0.80\columnwidth]{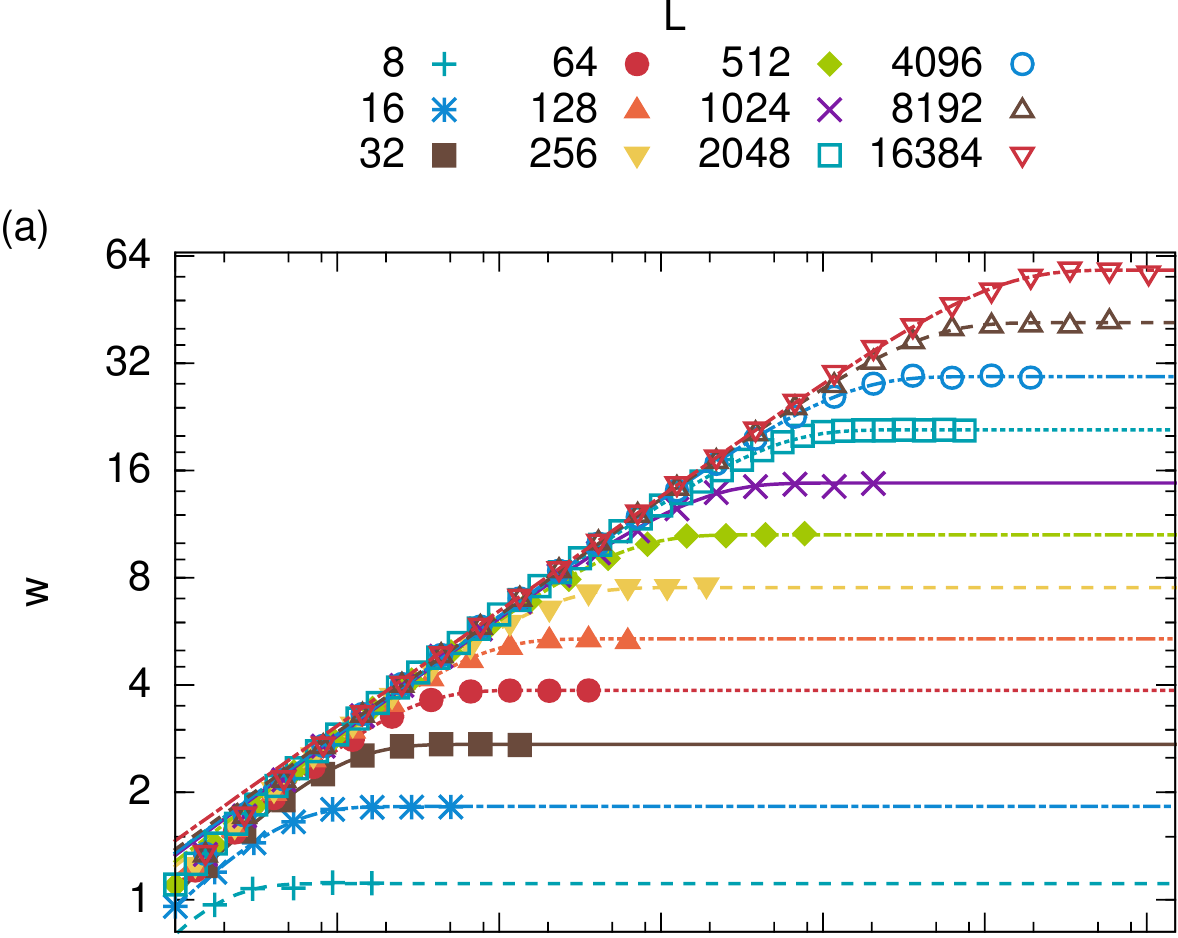}
\includegraphics[width=0.80\columnwidth]{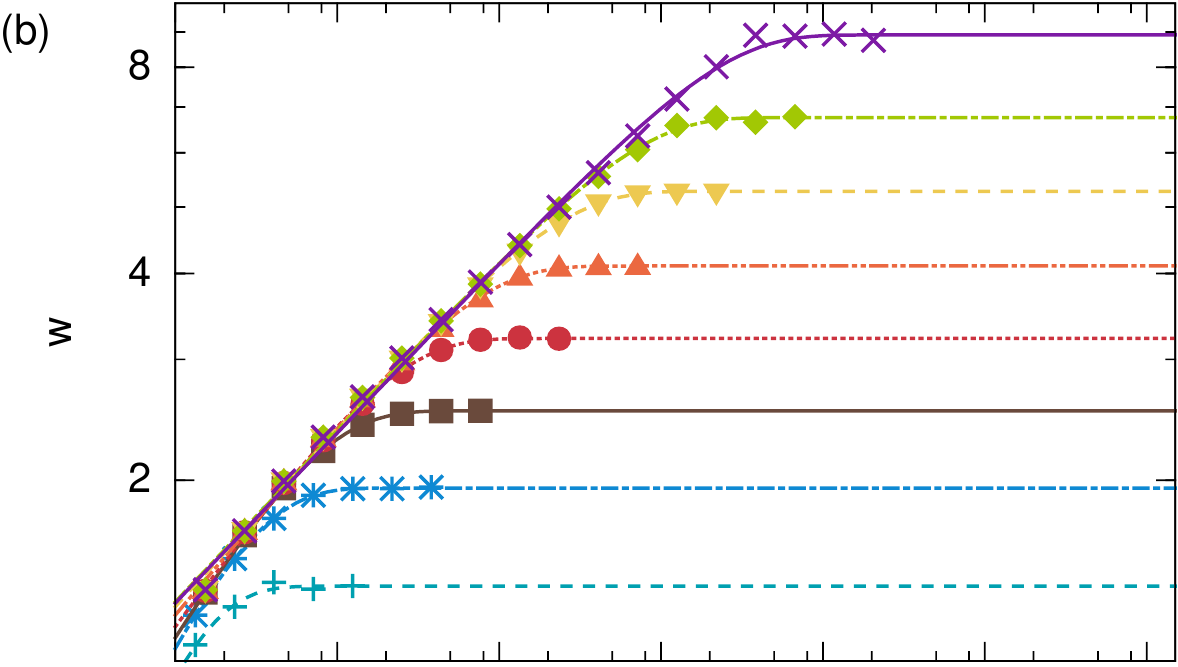}
\includegraphics[width=0.80\columnwidth]{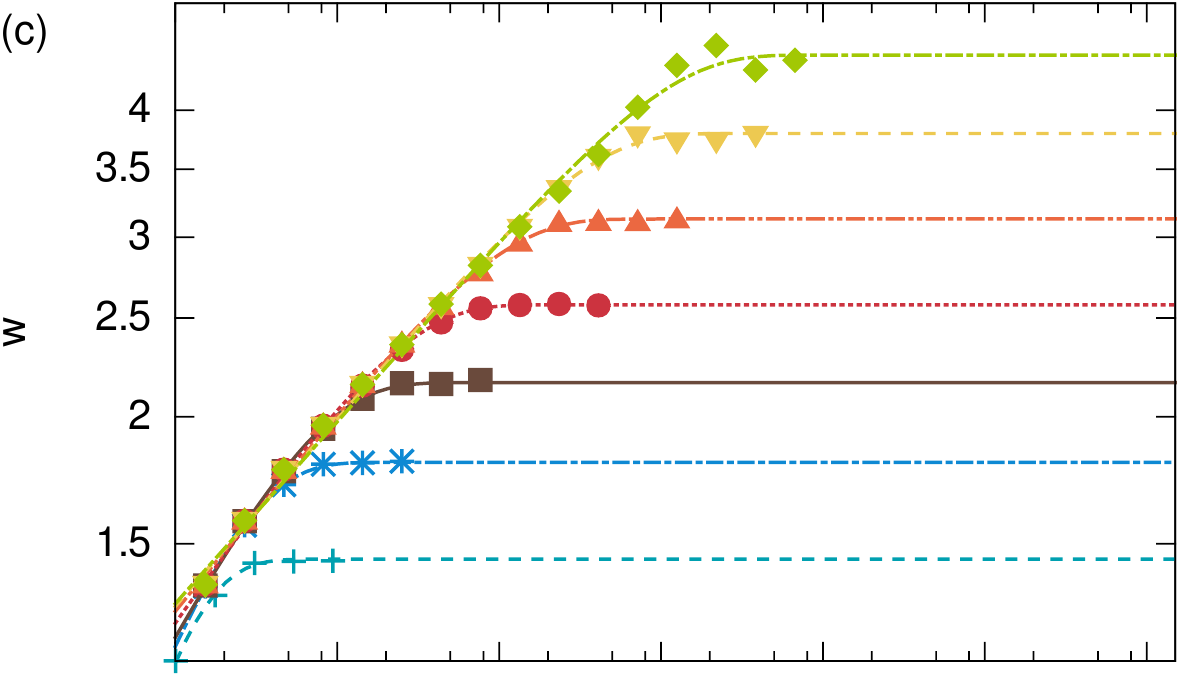}
\includegraphics[width=0.80\columnwidth]{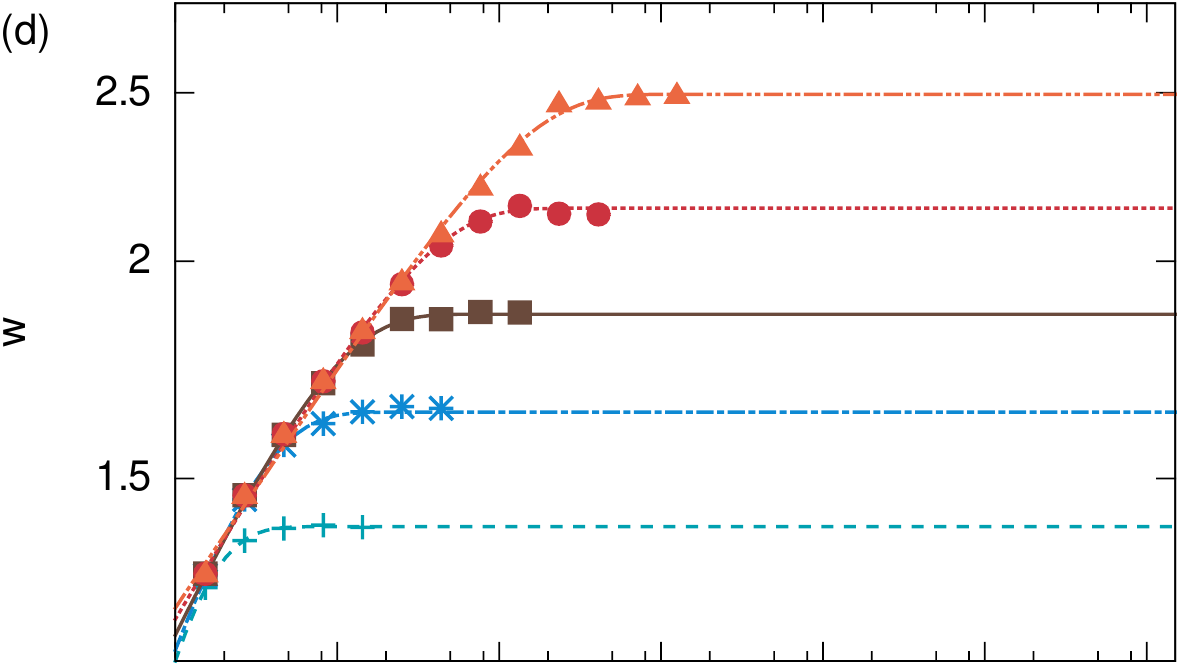}
\includegraphics[width=0.80\columnwidth]{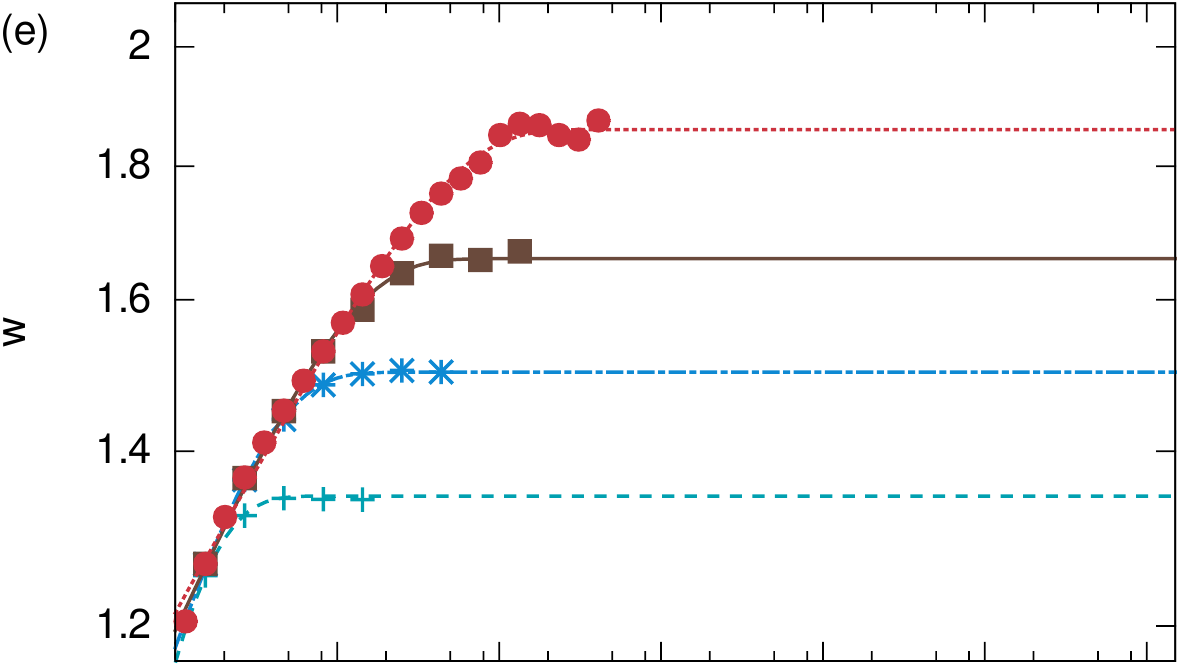}
\includegraphics[width=0.80\columnwidth]{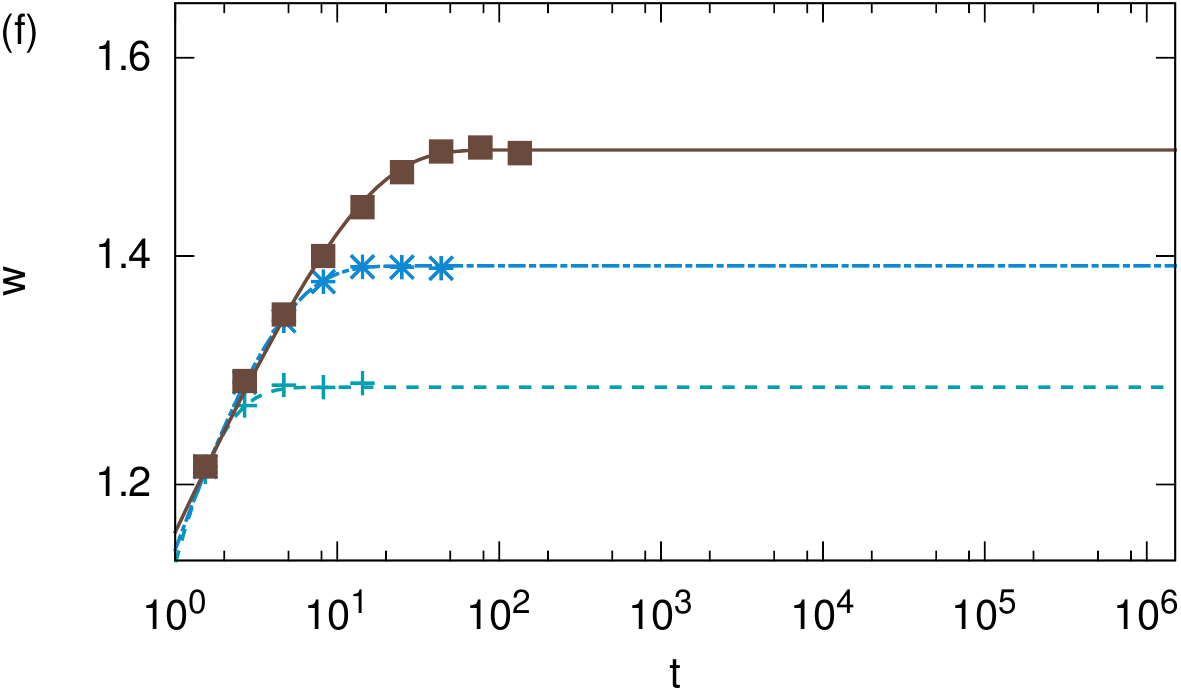}
\caption{
Roughness $w(L,t)$ from the etching model as a function of time, for surfaces with
(a) $d=1$, (b) $d=2$, (c) $d=3$, (d) $d=4$, (e) $d=5$, and (f) $d=6$, 
showing only data points for $t < 15 t_\times$ for
clarity. The lines are guides for illustration.
}
\label{fig:dynamics15d}
\end{figure}

We simulate several substrate lengths $L$ for each dimension $d$, with each experiment being
repeated several times. This ensemble average is necessary to reduce noise, producing
higher accuracy in the resulting exponents.

On figure \ref{fig:dynamics15d} we show our results for simulating roughness dynamics on
various substrate lengths for dimensions from $1+1$ to $6+1$, on log-log scale.
On all simulated dimensions, the expected Family-Vicsek  (FV) scalling \cite{Family1985}
is visible.

The Family-Vicsek scaling is a relation that can be used to model surface rougthness
dynamics by considering it composed of two different regimes: one in which it grows in a
power-law like function of time , and after a saturation time $t_\times$, it saturates. The
values of the saturation roughness is related to the substrate length, with
\begin{equation}\label{eq:lalpha}
w_s \propto L^\alpha,
\end{equation}
These
properties are expressed in the FV relation:
\begin{equation}\label{eq:FV}
  w(t,L) = w_s f(t/t_\times,\beta)= \begin{cases}
    w_y t^\beta  & \text{if $t \ll t_\times$}\\
     w_s & \text{if $t \gg t_\times$}\\
  \end{cases}.
\end{equation}
\end{section}

\begin{section}{Determination of exponents value}

Using the FV relation, we obtain our exponents by fitting our data to a set
of power laws. We fit values of $w_y$, $\beta_L$, and
$w_s$ by using the two expressions of  (\ref{eq:FV}) at $t\ll t_\times$ and
$t\gg t_\times$. Determination of $t_\times$ is made by analysing the intersection
between the functions of the aforementioned regimes.

The etching model presents a transient time at the beginning of the growth process,
and as such data from these times where $t\lesssim 1$ are discarded from the fitting.
This implies in a $\beta$ that is not independent of $L$. For this reason, the parameter
obtained from the fitting is called $\beta_L$ and a correction is made, in the form:
\begin{equation}
\beta_L = \beta \left(1+\frac{A_0}{L^\gamma}\right), \label{beta}
\end{equation}
where $\gamma \approx 1$.

This correction considers the real value of $\beta$ to be the asymptotic of $\beta_L$, 
eliminating finite size effects.

The values of $\beta_L$, $w_s$ and $t_\times$ for each value of $d$ and $L$
were obtained from roughness fitting and plotted in figure
\ref{fig:dim_ws_tx_beta}.

\begin{figure}
\includegraphics[width=0.8\columnwidth]{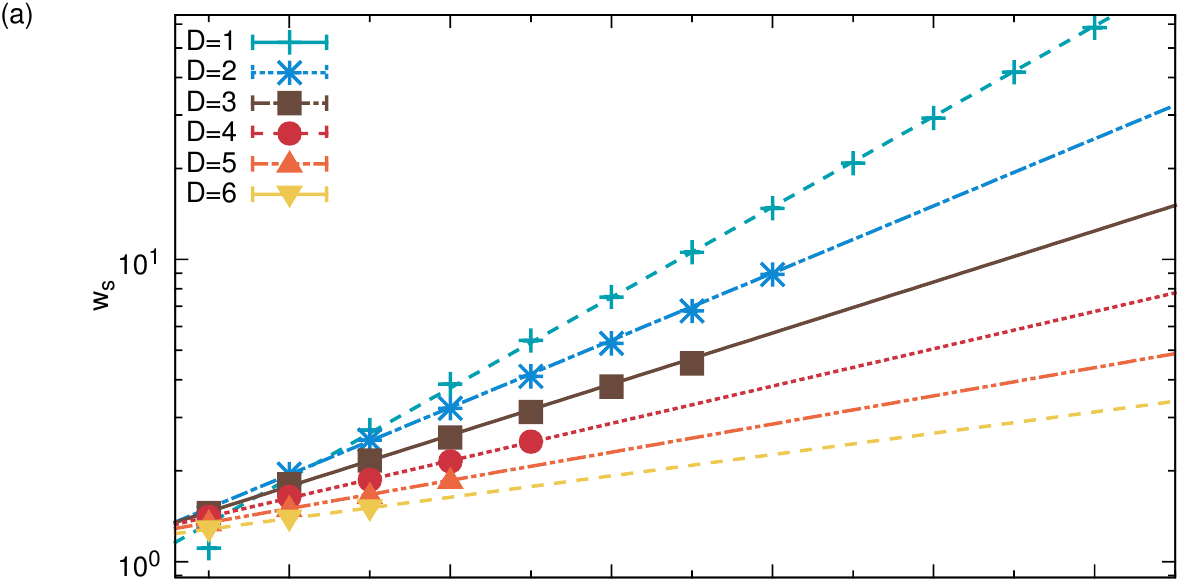}
\includegraphics[width=0.8\columnwidth]{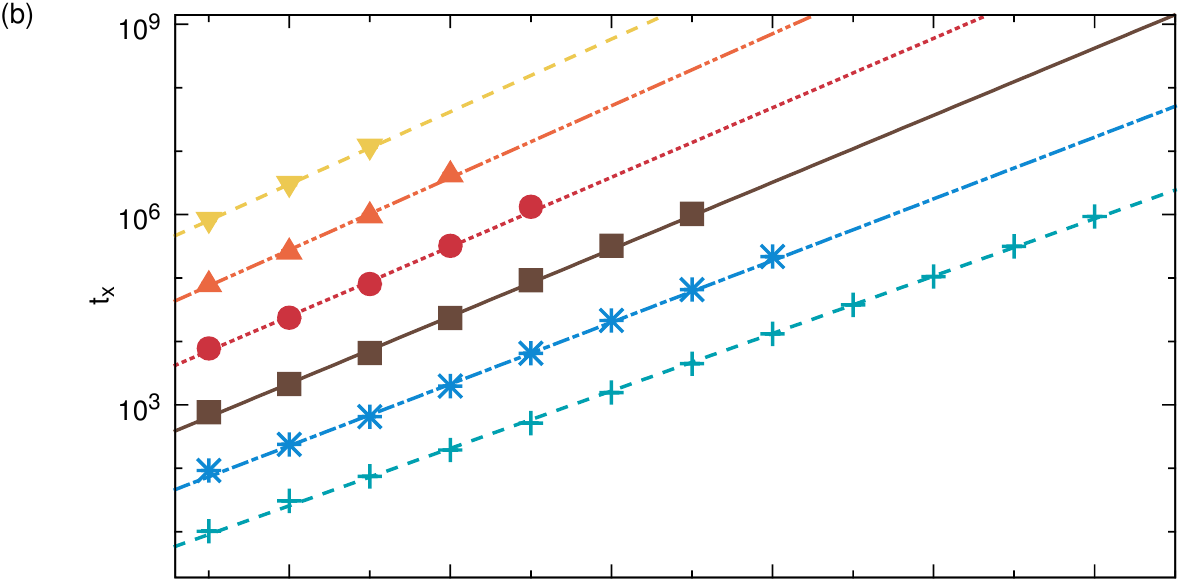}
\includegraphics[width=0.8\columnwidth]{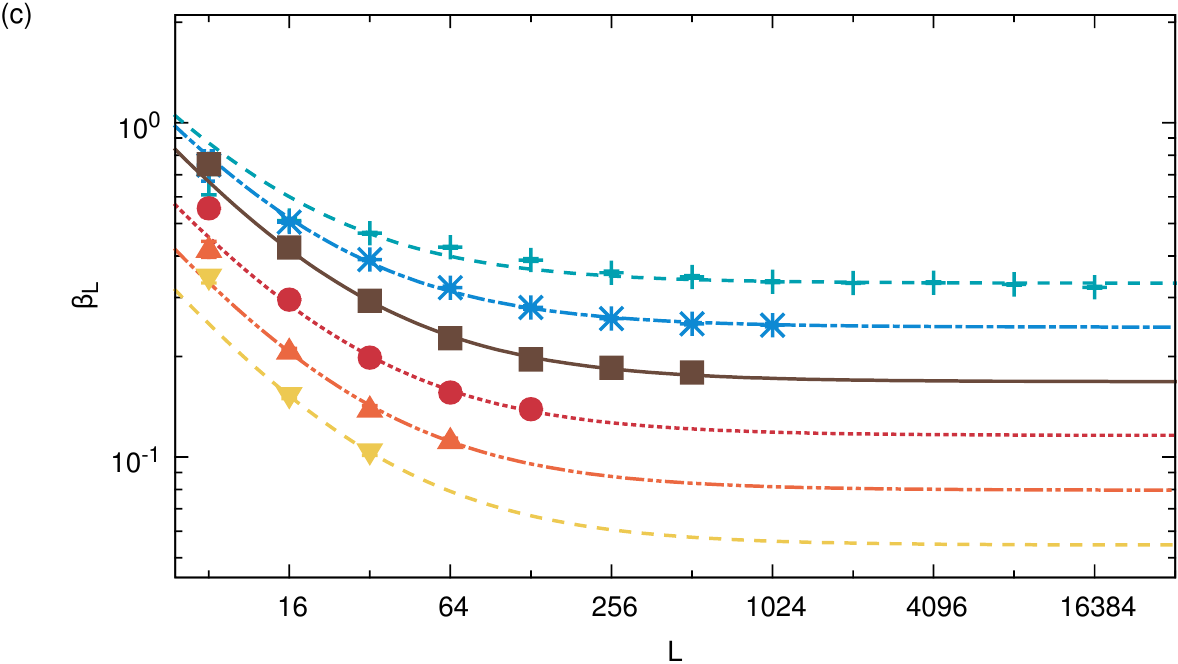}
\caption{Parameters of (\ref{eq:FV}) plotted as functions of $L$ with $d=1\dots6$.
(a) $w_s(L)$, (b) $t_\times(L)$ multiplied by $10^d$ for better visualization,
and (c) $\beta_L(L)$. In all simulations $L=2^n$, with $n$ integer.}
\label{fig:dim_ws_tx_beta}
\end{figure}

\end{section}

\begin{section}{Dynamic Exponents and the UCD}

Using the fitting from the data shown on figure \ref{fig:dim_ws_tx_beta}, we
obtain values for each exponent for dimensions ranging from $1+1$ to $6+1$. 
It allows us to observe how these expoents behave on higher dimensions. 

On table \ref{table:dfg_results} we show our results. It is simple to observe that
the expected behaviour for a system with a UCD on $d_c = 4$ is nowhere to be found,
with all exponents continualy changing. 
This suggests that there is not an UCD $d_c \le 6$, in agreement with previous results
\cite{Moser1991, Marinari2000, Marinari2002,Schwartz2012, Alves2014} obtained through
other models.

\begin{table}
\caption{Dynamic exponents obtained from the fittings of figure
\ref{fig:dim_ws_tx_beta}. A evidence of the precision of these exponents is the 
value of $\alpha+z$, which should be 2. 
\label{table:dfg_results}}
\centering
\begin{tabular}{ c c c c c } 
\hline
$d$           &$\alpha  $  &$\beta   $  &$z$          &$\alpha+z$ \\
\hline
\hline
$1$     & $0.497(5)$    &$0.331(3)$     &$1.50(8)$    & $2.00(1)$ \\
$2$     & $0.369(8)$    &$0.244(2)$     &$1.61(5)$    & $1.98(2)$ \\
$3$     & $0.280(7)$    &$0.168(1)$     &$1.75(9)$    & $2.03(2)$ \\
$4$     & $0.205(3)$    &$0.116(3)$     &$1.81(3)$    & $2.02(1)$ \\
$5$     & $0.154(2)$    &$0.079(3)$     &$1.88(6)$    & $2.04(1)$ \\
$6$     & $0.117(1)$    &$0.054(1)$     &$1.90(6)$    & $2.01(1)$ \\
\end{tabular}
\end{table}

It is important to note that although the one dimensional Etching model is on the KPZ
universality class, it is hard to classify it, or for that matter, any multi dimensional
model, on the KPZ universality class, as there are no known solutions for this case. We
can, however, compare our results with with others \cite{Rodrigues2015}, showing great 
concordance.

\end{section}

\begin{section}{Conclusion}

We have made a generalized version of the etching model, capable of simulating surfaces on
$d+1$ dimensions. Using this version of the model, we were capable of obtaining exponents for
systems up to $6+1$ dimensions.

Throught the obtained exponents, we have shown that the etching model does not show a upper critical
dimension at $d_c = 4$. It is not possible, however, to assert that the same thing is true
to the KPZ equation, as we still do not have a formal mapping of our model
to the KPZ equation, although comparison with current literature hints into the same
general direction of both the etching model belonging to the KPZ universality class
and the absence of this UCD on the KPZ equation.

We expect as well that importa results obtained in stochastic process 
\cite{Lapas2007,Morgado2007,Forgerini2009,Morgado2010,Ferreira2012,Ferreira2013,Fulinski2013} could
be used to give a more complete solution to this problem.

\end{section}

\begin{section}{Acknowledgements}
This work was supported by the Conselho Nacional de Desenvolvimento Cientifico
e Tecnologico (CNPQ), the Coordena\c{c}\~ao de Aperfei\c{c}oamento de Pessoal
de N\'ivel Superior (CAPES), the Funda\c{c}\~ao de Apoio a Pesquisa do Distrito
Federal (FAPDF), and the Companhia Nacional de Abastecimento (CONAB).
\end{section}


\section*{References}
\begin{thebibliography}{10}

\bibitem{Mello2001}
B.~A. Mello, A.~Chaves, and F.~A. Oliveira.
\newblock {\em Phys. Rev. E}, 63(4):041113, March 2001.

\bibitem{Kardar1986}
M.~Kardar, G.~Parisi, and Y.C. Zhang.
\newblock {\em Phys. Rev. Lett.}, 56(9):889--892, 1986.

\bibitem{Provatas1995}
N.~Provatas, T.~Ala-Nissila, M.~Grant, and K.~R. Elder.
\newblock {\em Phys. Rev. E}, 51(5):4232--4237, 1995.

\bibitem{Campos2013}
A.~T.~Campos and T.~M.~{da Rocha Filho}.
\newblock {\em Phys. A},
  392(18):3903--3908, September 2013.

\bibitem{Lita2000}
A.E. Lita and J.E. {Sanchez Jr}.
\newblock {\em Phys. Rev. B}, 61(11):7692, 2000.

\bibitem{Hwa1991}
T.~Hwa and E.~Frey.
\newblock {\em Phys. Rev. A}, 44(12):7873--7876, 1991.

\bibitem{Frey1996}
E.~Frey, U.~C T\"{a}uber, and T.~Hwa.
\newblock {\em Phys. Rev. E}, 53(5):4424--4438, May 1996.

\bibitem{Wio2013}
H.S. Wio, R.R. Deza, J.a. Revelli, and C.~Escudero.
\newblock {\em Acta Phys. Pol. B}, 44(5):889, 2013.

\bibitem{Katzav2002}
E.~Katzav and M.~Schwartz.
\newblock {\em Phys. A},
  309(1-2):69--78, June 2002.

\bibitem{Kim1991}
J.~M. Kim, J.M. Kosterlitz, and T.~Ala-Nissila.
\newblock {\em J. Phys. A}, 24:5569, 1991.

\bibitem{Moser1991}
K.~Moser, J.~Kert\'{e}sz, and D.E. Wolf.
\newblock {\em Phys. A},  178(2):215--226, 1991.

\bibitem{Stepanow1997}
S.~Stepanow.
\newblock {\em Phys. Rev. E}, 55(5):R4853--R4856, May 1997.

\bibitem{Lassig1998}
M.~L\"{a}ssig.
\newblock {\em Phys. Rev. Lett.}, 80(11):2366--2369, March 1998.

\bibitem{Bouchaud1993}
J.~P.~Bouchaud and M.~E.~Cates.
\newblock {\em Phys. Rev. E}, 47(3):1455--1458, 1993.

\bibitem{Forrest1990}
B.~M. Forrest and L.~H. Tang.
\newblock {\em Phys. Rev. Lett.}, 64(12):1405--1408, 1990.

\bibitem{Tang1992}
L.~H. Tang, B.~M. Forrest, and D.~E. Wolf.
\newblock {\em Phys. Rev. A}, 45(10):7162--7179, 1992.

\bibitem{Ala-Nissila1993}
T.~Ala-Nissila, T.~Hjelt, J.~M. Kosterlitz, and O.~Ven\"{a}l\"{a}inen.
\newblock {\em J. Stat. Phys.}, 72(1-2):207--225, 1993.

\bibitem{Ala-Nissila1998}
T.~Ala-Nissila.
\newblock {\em Phys. Rev. Lett.}, 80(4):9007, 1998.

\bibitem{Halpin-Healy1989}
T.~Halpin-Healy.
\newblock {\em Phys. Rev. Lett.}, 62(4):442--445, January 1989.

\bibitem{Wiese1998}
K.~J. Wiese.
\newblock {\em J. Stat. Phys.}, 93(1/2):143--154, October 1998.

\bibitem{Bhattacharjee1998}
J.~K. Bhattacharjee.
\newblock {\em J. Phys. A}, 31:L93--L96,
  1998.

\bibitem{Lassig1995}
M.~L\"{a}ssig.
\newblock {\em Nucl. Phys. B}, 50:559--574, 1995.

\bibitem{Lassig1997}
M.~L\"{a}ssig and H.~Kinzelbach.
\newblock {\em Phys. Rev. Lett.}, 85(5 Pt 1):050103, May 1997.

\bibitem{T.Blum1995}
A.~J.~McKane {T. Blum}.
\newblock {\em Phys. Rev. E}, 52(5):4741--4744, 1995.

\bibitem{M.A.MooreT.BlumJ.P.Doherty1995}
M.~A. Moore, T.~Blum, J.~P. Doherty, and M.~Marsili.
\newblock {\em Phys. Rev. Lett.}, 74(21):4257--4260, 1995.

\bibitem{Fogedby2009}
H.~C.~Fogedby.
\newblock {\em Pramana}, 71(2):253--262, February 2008.

\bibitem{Marinari2000}
E.~Marinari, A.~Pagnani, and G.~Parisi.
\newblock {\em J. Phys. A}, 33(2):8181, 2000.

\bibitem{Marinari2002}
E.~Marinari, A.~Pagnani, G.~Parisi, and Z.~R\'{a}cz.
\newblock {\em Phys. Rev. E}, 65(2):026136, January 2002.

\bibitem{Schwartz2012}
M.~Schwartz and E.~Perlsman.
\newblock {\em Phys. Rev. E}, 85(5):050103, May 2012.

\bibitem{Mello2015}
B.~A.~Mello.
\newblock {\em Phys. A}, 419:762--767, February 2015.

\bibitem{Rodrigues2015}
E.~A.~ Rodrigues, B.~A. Mello, and F.~A.~ Oliveira.
\newblock {\em J. Phys. A}, 48(3):035001, January 2015.

\bibitem{AaraoReis2006a}
F.~D.~A. {Aar\~{a}o Reis}.
\newblock {\em Phys A}, 364:190--196, May 2006.

\bibitem{Paiva2007}
T.~Paiva and F.~D.~A. {Aar\~{a}o Reis}.
\newblock {\em Surf. Sci.}, 601(2):419--424, January 2007.

\bibitem{Kimiagar2008}
S.~Kimiagar, G.~R. Jafari, and M.~R.~R. Tabar.
\newblock {\em J. Stat. Phys.}, 2008(02):P02010, February 2008.

\bibitem{Oliveira2008}
T.~J. Oliveira and F.~D.~A. {Aar\~{a}o Reis}.
\newblock {\em Phys. Rev. E}, 77:041605, 2008.

\bibitem{Tang2010}
G.~Tang, Z.~Xun, R.~Wen, K.~Han, H.~Xia, D.~Hao, W.~Zhou, X.~Yang, and Y.~Chen.
\newblock {\em Phys. A},
  389(21):4552--4557, November 2010.

\bibitem{Xun2012}
Z.~Xun, Y.~Zhang, Y.~Li, H.~Xia, D.~Hao, and G.~Tang.
\newblock {\em J. Stat. Mech.}, 2012(10):P10014, October 2012.

\bibitem{Family1985}
T.~{Family, F. Vicsek}.
\newblock {\em J. Phys A}, 18:L75, 1985.

\bibitem{Alves2014}
S.~G. Alves, T.~J. Oliveira, and S.~C. Ferreira.
\newblock {\em Phys. Rev. E}, 90(2):020103, August 2014.

\bibitem{Lapas2007}
L.~C Lapas, I.~V.~L Costa, M.~H Vainstein, and F.~a Oliveira.
\newblock {\em Europhys. Lett,}, 77(3):37004, February 2007.

\bibitem{Morgado2007}
R.~Morgado and M.~Cie\'{s}la.
\newblock {\em Europhys. Lett,}, 79(1):10002, July 2007.

\bibitem{Forgerini2009}
F.~L~Forgerini and W.~Figueiredo.
\newblock {\em Phys. Rev. E}, 79(4):041602, 2009.

\bibitem{Morgado2010}
C.~C.~Y. Dorea, J.~A.~Guerra, R.~Morgado, and A.~G.~C. Pereira.
\newblock {\em Numerical Functional Analysis and Optimization}, 31(2):164--171,
  2010.

\bibitem{Ferreira2012}
R.~M.~S. Ferreira, M.~V.~S. Santos, C.~C. Donato, J.~S. {Andrade Jr}, and F.~A
  Oliveira.
\newblock {\em Phys. Rev. E}, 86(2):021121, August 2012.

\bibitem{Ferreira2013}
R.~M.~S.~ Ferreira, L.~C.~ Lapas, and F~.A.~ Oliveira.
\newblock {\em Acta Physis. Polo. B}, 44(5):1085, 2013.

\bibitem{Fulinski2013}
A.~Fuliński.
\newblock {\em Acta Phys. Pol. B}, 44(5):1137, 2013.

\end{thebibliography}
\end{document}